\documentclass[12pt]{spieman}
\usepackage[T1]{fontenc}
\usepackage[utf8]{inputenc}
\usepackage{microtype}
\usepackage{amsmath,amsfonts,amssymb}
\usepackage{booktabs}
\usepackage{array}
\usepackage{graphicx}
\usepackage{enumitem}

\usepackage{xcolor}

\newcommand{\pxv}{PixelXVision}
\newcommand{\NR}{NR}

\title{An extensible software platform for automated operational and characterization testing of scientific cameras}
\author[a]{Baolong Chen}
\author[a,b,d*]{Hongfei Zhang}
\author[c]{Wei Liu}
\author[c]{Zihao Xia}
\author[b]{Xin Luo}
\author[a]{Qi Feng}
\author[a]{Zhe Geng}
\author[a,b,d*]{Jian Wang}
\affil[a]{School of Physical Sciences,
University of Science and Technology of China,
Hefei 230026, China}
\affil[b]{Institute of Advanced Technology,
University of Science and Technology of China,
Hefei 230031, China}
\affil[c]{Anhui GuoKe Vision PixelX Optoelectronics Technology Co., Ltd.,
Hefei 230031, China}
\affil[d]{Deep Space Exploration Laboratory,
Hefei 230088, China}
\begin{document}
\maketitle

\begin{abstract}
Multi-detector programs such as Earth 2.0 (ET) require detector characterization across different technologies, operating conditions, and production batches, together with consistent procedures for operation, characterization, and device selection. Traditional laboratory workflows often distribute camera control, illumination, image analysis, and report generation across detector-specific programs and manual steps, slowing iterative development and hindering standardized characterization. We address this gap with PixelXVision, a layered software platform that combines a common camera SDK, preset-based configuration, and a runtime plugin system. Detector-specific behavior is configured through declarative presets, allowing the same host interface to operate CCD, CMOS, and infrared detectors without model-specific core changes. Characterization procedures are implemented as plugins that use a shared object manager to access camera, illumination, storage, and database services. Each run links raw frames, measured operating conditions, configuration, analysis products, and reports into a traceable archive with a queryable index. The platform supports lightweight deployment for iterative single-detector development and structured workflows for multi-detector campaigns. The platform has been applied in characterization testing across CMOS, CCD, and HgCdTe cameras, and its configurable architecture can accommodate additional camera models for subsequent CCD testing for WFST without model-specific changes to the core host software. For operational troubleshooting, the platform integrates ChatPixel, a knowledge-base assistant that provides answers with cited sources when it has relevant information.
\end{abstract}


{\noindent\footnotesize
\textbf{*}Corresponding authors: Hongfei Zhang,
\href{mailto:nghong@ustc.edu.cn}{nghong@ustc.edu.cn};
Jian Wang,
\href{mailto:wangjian@ustc.edu.cn}{wangjian@ustc.edu.cn}.}

\begin{spacing}{2}

\section{Introduction}

Mosaic-camera programs like ET must characterize and select multiple large-format, high-sensitivity detectors before integrating them into the focal plane. These detectors often span different technologies, production batches, and operating conditions \cite{ge2022earth20}. Comparable detector-characterization requirements also arise in projects such as the Wide Field Survey Telescope (WFST) \cite{wang2023wfst}, where detector testing and evaluation must accommodate evolving devices and operating conditions. More generally, scientific detector characterization may involve CCD, CMOS, and infrared devices with substantially different readout architectures, operating temperatures, response characteristics, and noise behavior. Standard characterization commonly reports linearity and dark-current behavior \cite{yu2018kasi,widenhorn2002temperature}, together with read noise, conversion gain, full-well capacity, and defective-pixel statistics \cite{janesick1987photon,rosenberger2016emva1288}. Radiometric camera-calibration methods provide a complementary basis for interpreting response and noise estimates \cite{healey1994radiometric}. Repeatable and reliable characterization workflows must record the measured operating conditions for each run and link them to the acquired data across devices and test sessions. This is especially true during development and debugging, when characterization often runs on a single camera or a small lab setup, and acquisition parameters may change.

Most existing characterization pipelines—whether built for a specific instrument or for a reconfigurable testbed—split control, acquisition, and analysis into separate, device-specific layers. This works well enough for the target configuration, but becomes a maintenance burden when the hardware revision, readout mode, or detector type changes. The provenance linking raw frames to operating conditions and analysis steps is often left implicit \cite{simmhan2005provenance,kazic2015provenance}. Operators then manually track device state, acquisition order, parameter settings, and measurement conditions. The burden grows for cameras still under commissioning: hardware revisions and acquisition-mode updates can change the effective test environment without warning. In the test for ET, frequent short exposures caused the detector temperature to drift away from the cooling setpoint during acquisition, and this affected dark current. If these conditions are not recorded and linked to the frame data, subsequent analysis proceeds as if the data were taken under nominal settings. Since some characterization depends on temperature, this mismatch can bias the estimates and reduce comparability across characterization runs \cite{widenhorn2002temperature,rieke2015miri,khan2025ccd}.

Existing systems address different parts of detector characterization. Mission-specific efforts include detector characterization for Euclid NISP, focal-plane testing for LSST, and characterization of the Roman H4RG-10 detectors \cite{secroun2016euclidh2rg,snyder2020lsst,mosby2020romanh4rg}. Reconfigurable laboratory testbeds support VIS--NIR--SWIR detector testing and precision CCD characterization \cite{schindler2014ingaasbench,khan2025ccd,shapiro2019ppl}. General-purpose control frameworks such as EPICS, TANGO, ACS, and INDI provide reusable infrastructure for instrument control \cite{dalesio1994epics,gotz2003tango,chiozzi2004acs,hoffmann2022indi}, while orchestration systems such as RTS2, pyobs, and Bluesky support programmable execution and metadata management \cite{kubanek2010rts2,husser2022pyobs,allan2019bluesky,zhao2023bluesky}. Table~\ref{tab:reported-capabilities} compares these and related systems using six criteria relevant to cross-detector characterization campaigns: reusable hardware interfaces, unified characterization across detector technologies, detector-specific workflows, run-context capture, automated reporting, and a self-contained run archive with an index. Among the prior systems listed in Table~\ref{tab:reported-capabilities}, different systems cover different subsets of these criteria; no prior system is reported to cover all six. PixelXVision combines all six capabilities in one platform and supports both single-detector development and multi-detector characterization campaigns.
\begin{table*}[t]
\centering
\caption{Reported architectural and workflow capabilities of representative systems.}
\label{tab:reported-capabilities}

\footnotesize
\setlength{\tabcolsep}{3pt}

\begin{tabular}{
    >{\raggedright\arraybackslash}p{0.25\linewidth}
    *{6}{>{\centering\arraybackslash}p{0.09\linewidth}}
}
\toprule
& \multicolumn{6}{c}{Reported capability criteria} \\
\cmidrule(lr){2-7}

System and scope
& \shortstack{Reusable\\interface}
& \shortstack{Unified\\cross-tech.\\char.}
& \shortstack{Detector\\workflow}
& \shortstack{Character.\\run context}
& \shortstack{Generated\\report}
& \shortstack{Self-cont.\\archive +\\index} \\

\midrule

Euclid/Roman qualification
\cite{secroun2016euclidh2rg,waczynski2016euclid,mosby2020romanh4rg}
& \NR & \NR & R & \NR & \NR & \NR \\

LSST focal-plane testing
\cite{snyder2020lsst}
& \NR & \NR & R & \NR & \NR & \NR \\

Schindler/Khan testbeds
\cite{schindler2014ingaasbench,khan2025ccd}
& \NR & \NR & R & \NR & \NR & \NR \\

Precision Projector Laboratory
\cite{shapiro2019ppl}
& \NR
& \NR\textsuperscript{a}
& R
& \NR
& \NR
& \NR \\

Common control frameworks (EPICS, TANGO, ACS, INDI)
\cite{dalesio1994epics,gotz2003tango,chiozzi2004acs,hoffmann2022indi}
& R & \NR & \NR & \NR & \NR & \NR \\

LSST CCD test suite (RTS2)
\cite{prouza2010rts2ccd,kubanek2010rts2}
& R
& \NR
& R
& R
& \NR
& \NR\textsuperscript{b} \\

pyobs
\cite{husser2022pyobs}
& R & \NR & \NR & R & \NR & \NR \\

Bluesky (RunEngine core)
\cite{allan2019bluesky,zhao2023bluesky}
& R & \NR & \NR & R & \NR & \NR \\

LIMA
\cite{limaReference}
& R & \NR & \NR & \NR & \NR & \NR \\

areaDetector
\cite{areadetectorReference}
& R & \NR & \NR & \NR & \NR & \NR \\

GenICam
\cite{mulholland2020genicam}
& R & \NR & \NR & \NR & \NR & \NR \\

Micro-Manager
\cite{stuurman2010micromanager}
& R & \NR & \NR & \NR & \NR & \NR \\

\pxv{} (this work)
& R & R & R & R & R & R \\

\bottomrule
\end{tabular}

\vspace{3pt}

\begin{minipage}{\textwidth}
\footnotesize
\raggedright
\textsuperscript{a}Configurable for optical and NIR sensors; cross-technology
software reuse was not explicitly evaluated.

\textsuperscript{b}Observation database and archive exist, but
characterization-specific run packaging and a self-contained,
rebuildable index were not reported.

\textit{Note:} \textit{R} denotes that the capability is reported in
the cited sources; \textit{NR} denotes that it is not reported in the
cited sources and does not necessarily imply that the capability is absent.
\end{minipage}

\end{table*}

What Table~\ref{tab:reported-capabilities} makes clear is that while each of these systems does one or several things well—control frameworks offer reusable interfaces, testbeds provide characterization workflows, and so on—none delivers an integrated platform that meets all six criteria at once. This is not an abstract concern. Multi-detector campaigns such as the ET irradiation program need a platform that supports both rapid single-camera iteration and structured multi-device scaling. They also require cross-technology comparison for detector selection, frequent procedure updates during debugging, and long-term reproducibility.

We address this gap with PixelXVision, a layered software platform built around a common camera SDK and declarative detector presets. The SDK provides a unified device-control interface for CCD, CMOS, and infrared cameras, while the host loads model-specific presets according to the device-reported identifier. These presets define parameters, controls, status monitoring, and validation rules. Once the corresponding SDK device implementation is available, additional cameras can be integrated without changes to the model-independent host core.

Characterization procedures are implemented as runtime plugins that access camera, light-source, storage, and database services through shared platform interfaces. The implemented measurements include dark current, read noise, photon-transfer curves, spatial nonuniformity, and defective-pixel detection, and can be combined into automated workflows. For each run, the platform associates raw FITS frames and measured operating conditions with the corresponding configuration, analysis products, and reports, while SQLite indexes run metadata and metrics for historical queries and reanalysis. PixelXVision also includes ChatPixel, a knowledge-retrieval assistant for camera operation and diagnostics. ChatPixel uses the device model, firmware version, error state, and runtime signals to guide retrieval from SDK and platform documentation, and returns a source-linked answer only when the retrieved evidence is sufficient.

The remainder of this paper is organized as follows. Section~\ref{sec:architecture} describes the platform architecture, camera abstraction, preset-based configuration, and plugin mechanisms. Section~\ref{sec:traceability} presents the run-data and traceability model. Section~\ref{sec:chatpixel} describes ChatPixel and its retrieval workflow. Section~\ref{sec:evaluation} evaluates the implemented workflows using CCD, CMOS, and infrared cameras.
\section{System Architecture}
\label{sec:architecture}

\subsection{Design goals}

\pxv{} supports scientific-camera testing throughout development, manufacturing, and operational use. During development, camera selection, readout electronics, communication interfaces, and operating parameters may change before a complete instrument-control system is available, but controlled, repeatable tests are required to guide design decisions. During manufacturing, the same tests must be repeated across devices for batch comparison and acceptance testing; after design maturation, they remain essential for calibration and periodic performance reassessment. The architecture therefore separates characterization workflows from individual camera implementations and keeps camera-control details outside the test procedures.

Building on this separation, the system is organized into three layers: camera control, platform, and characterization extensions. The camera-control layer is implemented by VPXSDK, which exposes a unified camera abstraction for camera discovery, parameter configuration, exposure and triggering, temperature control, region-of-interest selection, binning, and image readout. VPXSDK is a separate library and can be used independently of the upper-level \pxv{} workflows.

The platform layer combines the platform kernel and the host application. The kernel provides a runtime type system, object management, property-based state, event dispatch, persistent configuration, preset management, and plugin management. The host application handles the user interface, parameter validation, preset loading, run-state management, and camera-operation coordination, and maps settings and acquisition requests to VPXSDK through the shared \texttt{SDKParameters} object. Presets provide model-specific parameter mappings, interface definitions, and validation configurations, while the configuration service stores persistent user and plugin settings.

The characterization-extension layer consists of characterization plugins and shared service plugins. Characterization plugins implement the acquisition procedures, state control, and analysis algorithms of individual tests. They obtain camera capabilities through \texttt{SDKParameters} and other runtime resources through platform interfaces; they do not handle camera protocols or device communication directly. Shared service plugins provide functions such as light-source and shutter control, data storage, and database synchronization. For each run, the platform stores the camera identity, test parameters, acquired data, analysis results, and generated report.

\subsection{Layered organization}

\begin{figure}[t]
    \centering
    \includegraphics[width=0.96\linewidth]{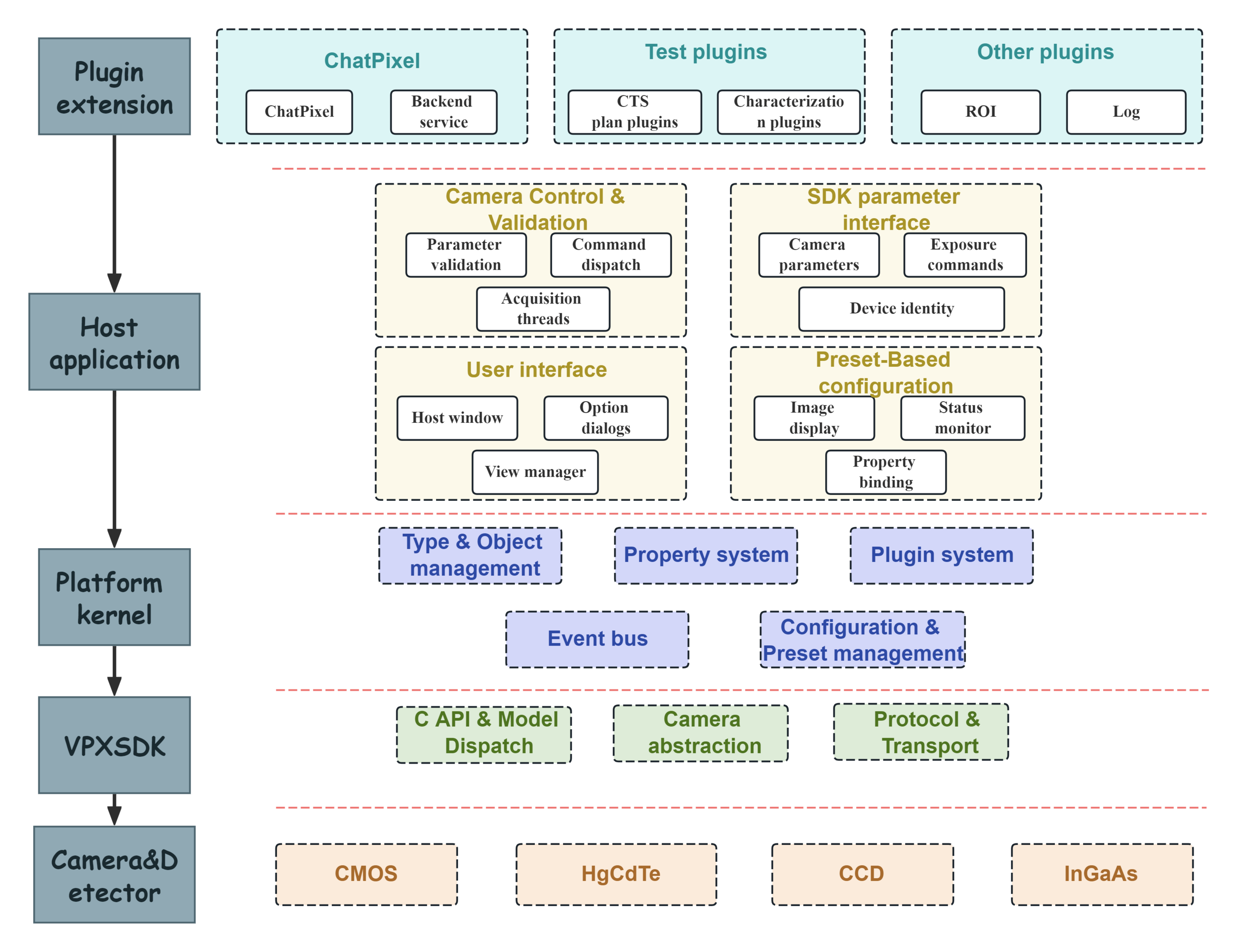}
    \caption{Layered organization of \pxv{} and the relationships among characterization extensions, the host application, platform kernel, VPXSDK, and physical cameras and detectors.}
    \label{fig:architecture}
\end{figure}

Figure~\ref{fig:architecture} shows the relationships among the characterization extensions, host application, platform kernel, VPXSDK, and physical cameras and detectors. VPXSDK lies between the platform and the hardware. Within VPXSDK, the camera API and model dispatch sit above the protocol and transport implementations.

Within the host application, camera initialization and parameter control follow a common sequence. After the user selects a connection type, VPXSDK enumerates the connected device and returns its model identifier. The host maps the identifier to the corresponding camera subtype and loads the matching XML preset. The host uses the preset to initialize parameter mappings, interface fields, monitoring settings, and validation rules. The shared \texttt{SDKParameters} object then exposes camera parameters and acquisition operations to the host and characterization plugins. For a requested parameter change, \texttt{VPXVerify} applies the preset-defined validation rules, invokes the corresponding VPXSDK operation, and updates the associated property after the operation succeeds. The resulting property change is published through the platform event mechanism to registered listeners.

\subsection{Platform kernel}

At startup, the base, user-interface, and application modules register their runtime types before creating shared objects such as \texttt{SDKParameters} and \texttt{Args}. Loaded plugins can extend the same runtime type system through their \texttt{registerType} hooks and create additional shared objects when required.

Runtime objects, including camera-parameter objects, images, regions of interest, and shared services, can be located by type and object identifier. The object-management service creates objects from registered runtime types and provides access to existing objects. Properties represent object state and are used for camera parameters, image metadata, and other runtime data.

The kernel distinguishes runtime state from persistent configuration. Object properties represent the current state of cameras, views, and workflows, while the configuration service stores persistent user and plugin settings. The preset-management service supplies the preset definitions the host uses during model-specific initialization. Camera-specific validation remains in the host-level \texttt{VPXVerify} component rather than in the kernel.

The event manager publishes property and configuration changes and dispatches them to registered listeners.

\begin{figure}[h]
    \centering
    \includegraphics[width=0.92\linewidth]{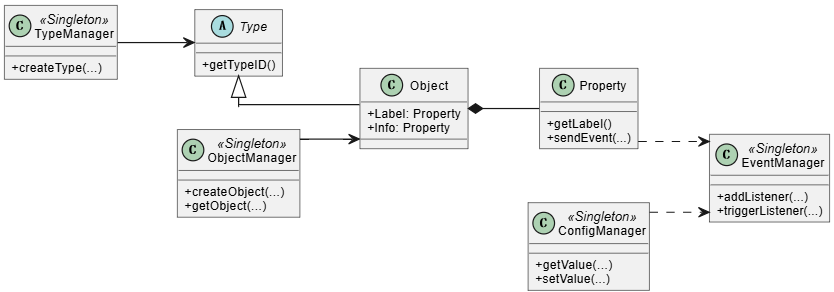}
    \caption{Selected class relationships for the runtime type system, object management, property-based state, configuration access, and event dispatch in the \pxv{} platform kernel.}
    \label{fig:kernel-classes}
\end{figure}

Figure~\ref{fig:kernel-classes} shows the principal class relationships for the runtime type system, object management, property-based state, configuration, and event dispatch. \texttt{TypeManager} maintains the runtime type system, \texttt{ObjectManager} creates and retrieves runtime objects, and \texttt{Object} derives from \texttt{Type} and associates state with \texttt{Property}. \texttt{Property} and \texttt{ConfigManager} use \texttt{EventManager} to publish change notifications. The figure omits plugin- and preset-management classes.

\subsection{Camera abstraction and preset-based configuration}

VPXSDK defines the camera-control boundary between the host application and device-specific implementations. Its external C API exposes common operations for exposure, triggering, temperature and fan control, gain, binning, region-of-interest selection, and image acquisition. At the same time, model dispatch selects the corresponding camera implementation. Internally, VPXSDK separates the camera abstraction from the protocol and transport layers.

At the class level, \texttt{VPXDevice} provides the device-level base, and the abstract \texttt{VPXCamera} class defines common camera-control behavior. CCD, CMOS, and HgCdTe camera families derive from \texttt{VPXCamera}, while \texttt{VPXVirtualCamera} provides a concrete virtual-camera implementation. Family-level and model-specific classes implement device-specific behavior.

\begin{figure}[h]
    \centering
    \includegraphics[width=0.92\linewidth]{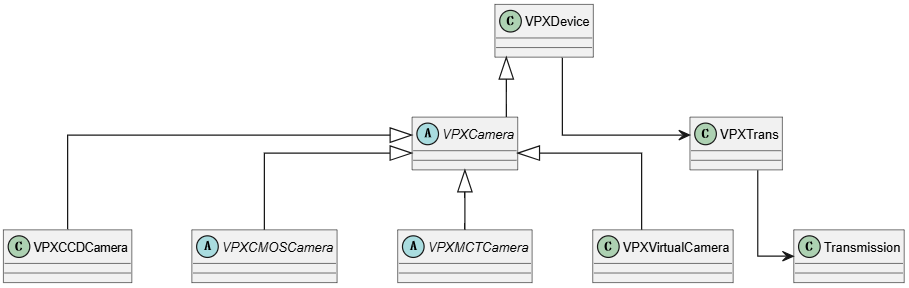}
    \caption{Simplified VPXSDK class relationships showing the camera abstraction, detector-family implementations, and protocol and transport path.}
    \label{fig:vpxsdk-classes}
\end{figure}

\texttt{VPXTrans} and \texttt{Transmission} implement the communication path. \texttt{VPXDevice} maintains a \texttt{VPXTrans} object for protocol-aware communication, while \texttt{VPXTrans} delegates the underlying I/O to a \texttt{Transmission} backend. Figure~\ref{fig:vpxsdk-classes} shows these class relationships.

The host and characterization plugins access VPXSDK through the shared \texttt{SDKParameters} object described above. A new camera model requires a corresponding VPXSDK device implementation and a model-specific preset. Once these are provided, the model-independent host core need not change.

\subsection{Plugin contract}

\pxv{} loads characterization workflows and shared services through a common plugin contract. An active plugin is distributed as a shared library with an XML manifest. At startup, the plugin manager scans the plugin directories, loads the shared library, resolves its \texttt{createPlugin} entry point, reads the manifest, and invokes the plugin's \texttt{registerType} hook.

A plugin can register additional runtime types and shared objects. Its manifest contains plugin metadata and may declare user-interface actions and persistent settings, which the host uses to construct the corresponding controls.

After loading, a plugin exposes workflow or service operations through the common \texttt{process} interface. Plugins retrieve shared platform objects through the object-management service, including \texttt{app::SDKParameters}, \texttt{app::Image}, and storage services. Persistent values are namespaced by plugin identifier within the common configuration subsystem.

The plugin manager handles installation and lifecycle states for packaged extensions. Changes to these states take effect after application restart. Characterization procedures such as dark-current, photon-transfer, and gain/read-noise measurements can be added as workflow plugins without touching the platform kernel. Support for a new camera model is handled separately through the corresponding VPXSDK device implementation and model-specific preset.
\section{Traceability and Output Model}
\label{sec:traceability}

The platform archives each completed test run as a self-contained directory containing the raw FITS frames, acquisition metadata, analysis products, figures, and report. A storage service registers the directory and updates the test and FITS indexes. In contrast, a separate database-synchronization service imports the metadata, results, and analysis metrics into a local SQLite database for historical queries. The run directory remains the authoritative archive; the file indexes and SQLite database are derived lookup layers.

FITS serves as the archival image container. The SDK and acquisition layer supply camera and acquisition fields, and the characterization plugin adds test-specific fields \cite{pence2010fits}. The plugin fields can include the test and image types, test sequence, exposure time, frame number, and measured temperatures. These fields capture each frame's acquisition context, while the run directory and its metadata link the frame to the corresponding test.

Archived runs can be inspected and reanalyzed without access to the live camera. Because each run directory retains the raw frames and acquisition metadata, the characterization plugin can be rerun on the same frames with revised settings — for example, a different ROI — without repeating the acquisition. The raw frames, acquisition metadata, analysis products, and reports remain associated with the same run during this process. This organization follows established provenance practices by retaining source data, acquisition context, and derived products together \cite{simmhan2005provenance,moreau2015prov,kazic2015provenance}.

\begin{figure}[t]
\centering
\includegraphics[width=\linewidth]{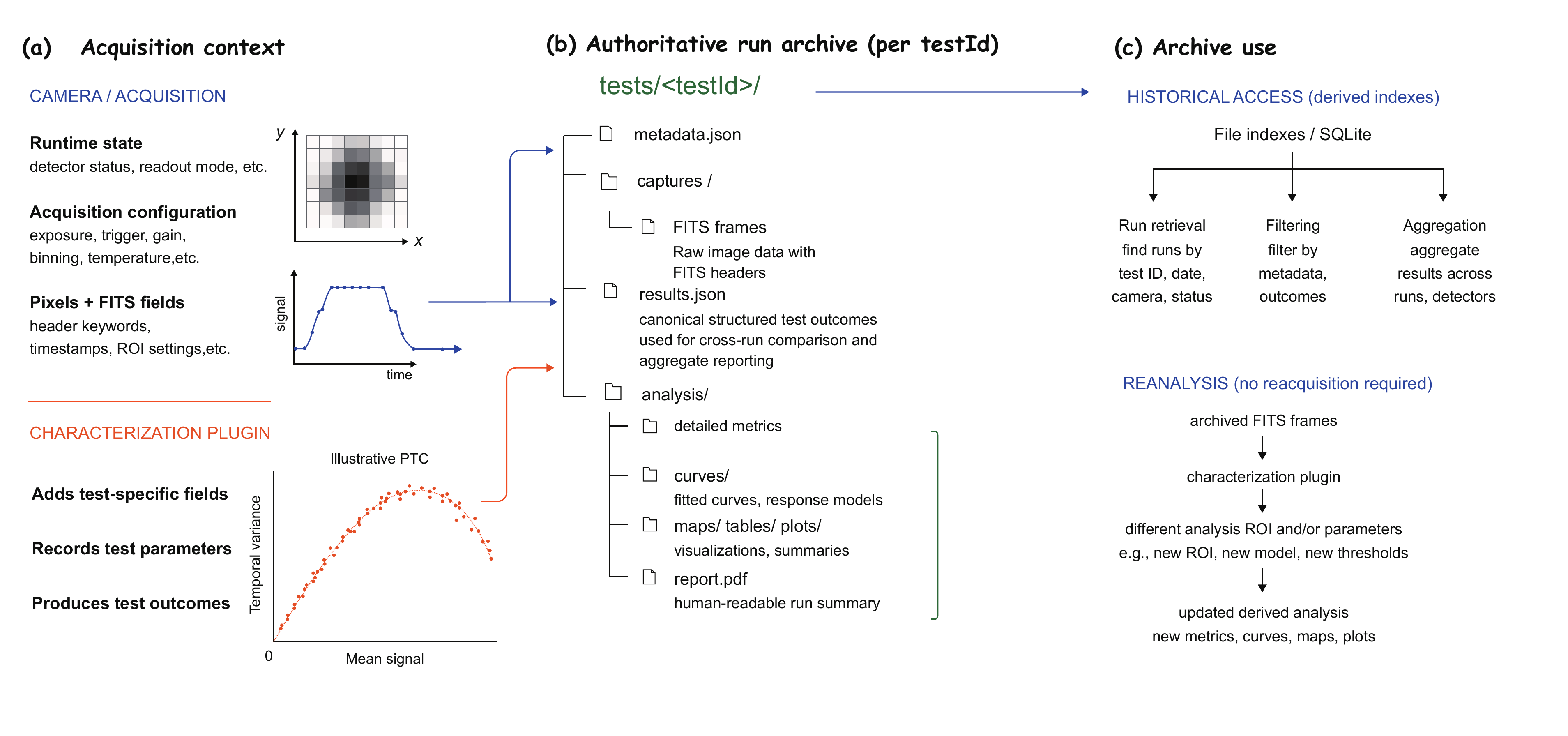}
\caption{Data flow for a single test run, showing the run directory, FITS data and metadata, SQLite indexing, and plugin-driven reanalysis.}
\label{fig:traceability}
\end{figure}

\section{Knowledge Retrieval Assistance}
\label{sec:chatpixel}

Scientific camera testing requires model-specific information from camera manuals, SDK interfaces, plugin documentation, error-code definitions, and test-method descriptions, along with automated acquisition and analysis. Operators consult these sources when configuring tests or interpreting runtime conditions. \pxv{} includes ChatPixel, a read-only knowledge-retrieval assistant for camera operation and characterization. ChatPixel follows a retrieval-augmented generation (RAG) architecture that combines an explicit technical knowledge base with platform context \cite{lewis2020rag}. The camera model, firmware version, error state, and selected runtime signals guide retrieval and query disambiguation. ChatPixel returns source-linked information but does not issue camera-control commands or modify test execution.

\subsection{Overall architecture}

ChatPixel is divided between a plugin in \pxv{} and an independently deployable retrieval backend. The plugin provides the user interface, reads current platform state through existing object and property interfaces, manages local settings, and sends queries to the backend. At query time, the plugin can attach a bounded, read-only context snapshot. Identification fields include the camera model, firmware version, and detector identifier. Error context includes current device and platform error codes and messages. Runtime context includes acquisition and thermal parameters, selected ROI and image metadata, image statistics, loaded plugins, platform state, and recent logs.

An explicit error code is retained for direct matching; if the query omits one, the current platform error state can provide it. Camera-model, firmware, and detector identifiers extend the retrieval query, while camera-characterization queries are routed to platform test-method sources. Because knowledge processing runs in the backend, it remains separate from the camera-acquisition runtime, and the knowledge base and retrieval logic can be updated without changing the platform plugin.

\begin{figure*}[ht]
\centering
\includegraphics[width=\textwidth]{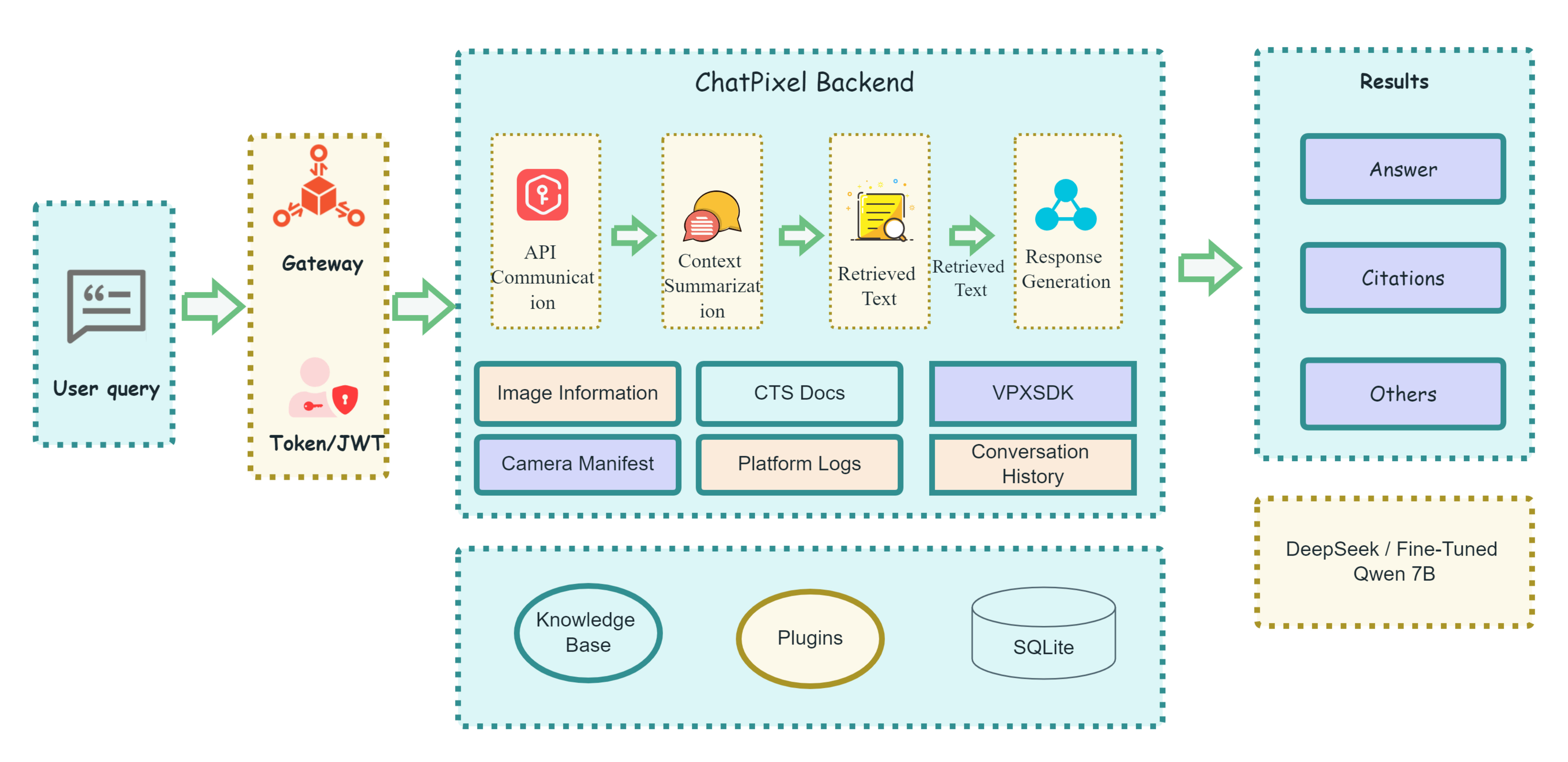}
\caption{ChatPixel architecture and query path from platform context to retrieval, answer generation, citation, and evidence-gated refusal.}
\label{fig:chatpixel-architecture}
\end{figure*}

Figure~\ref{fig:chatpixel-architecture} summarizes the query path from context-snapshot ingestion to retrieval, answer generation, citation, and evidence-gated refusal.

An explicit source manifest limits backend ingestion to selected platform documentation, CTS plugin descriptions, VPXSDK headers, error-code definitions, test-method documents, and camera-configuration files. Each source is divided into text chunks identified by a stable identifier, file path, title, section, and tags. During indexing, the backend also extracts structured records for error codes, API symbols and signatures, plugin manifests, recurring platform-log patterns, and characterization and test concepts with English, Chinese, and pinyin aliases. At query time, it identifies intents, numeric or symbolic error codes, SDK API names, test concepts, and source groups. The backend then combines structured matches with lexical retrieval through SQLite FTS5 and uses BM25 to rank document candidates \cite{robertson2009bm25}. Exact SDK symbols, error codes, and log fragments are matched to structured records, whereas general operational questions are retrieved from document chunks. The system routes model-specific test questions to curated test-method sources, and both structured records and document chunks retain links to their sources for citation.

After normalizing the question and attached context, the backend constructs a retrieval query and ranks the available sources. It rewrites a short follow-up only when the query has low confidence and the recent conversation supplies its missing topic, following conversational query-rewriting methods \cite{yu2020queryrewriting}. Numerical identifiers, error codes, API symbols, and log fragments remain unchanged during rewriting. Questions about current temperature, image statistics, image metadata, cooling state, and the latest log use the read-only context, whereas explanations of errors, APIs, plugins, and test methods require retrieved evidence. With a configured language model, the backend organizes the evidence into a concise answer; otherwise, it returns a deterministic extractive summary. Knowledge-based answers include the source path, section, and supporting snippet, while the response records its generation mode and query plan \cite{gao2023citations}. If retrieval yields no sufficiently reliable source, the backend returns no supported answer rather than an unsourced operational instruction. This behavior implements an evidence gate for knowledge-based operational answers \cite{ji2023hallucination,rajpurkar2018unanswerable}.

\section{Implementation Evaluation}
\label{sec:evaluation}

We evaluated the \pxv{} workflow on three cameras with different detector types: the VPX1081 CMOS, the PixelX VPXIR-MCT20, and the CCD4720. The workflow and results presented in this evaluation are based on actual camera tests performed using the platform and the corresponding platform-generated reports. The evaluation focused on three aspects: model-specific device integration, execution through the common host and plugin mechanisms, and generation of run-linked data products.
\subsection{Camera and detector coverage}

\begin{figure}[ht]
    \centering
    \includegraphics[width=\linewidth]{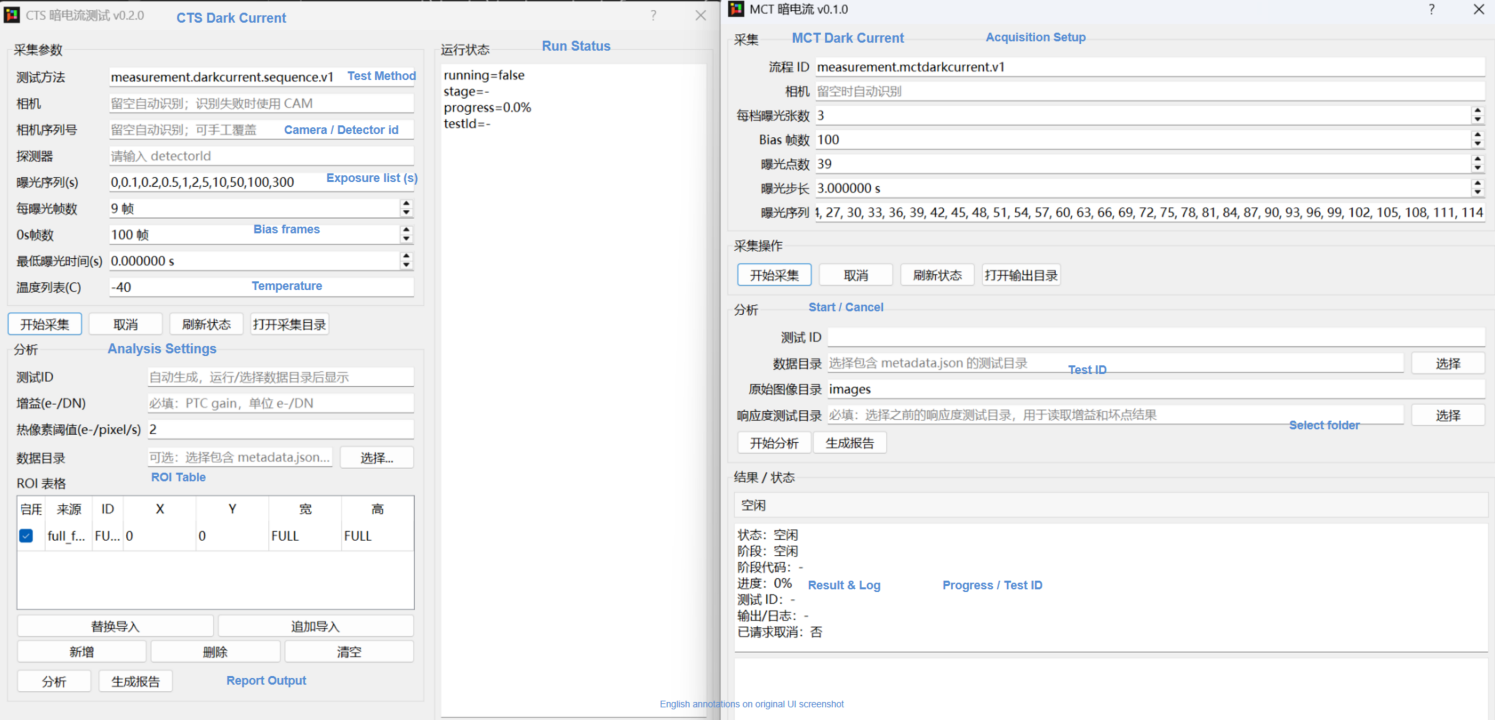}
    \caption{Plugin-based configuration interfaces for the CTS (left) and MCT (right) dark-current tests. The CTS panel provides flexible parameter control for research use, while the MCT panel adopts a simplified layout for batch testing. Both panels share the same workflow structure and support multiple detector types.}
    \label{fig:plugin-ui}
\end{figure}

Each camera model has a corresponding VPXSDK device class, USB or fiber transport configuration, and model-specific preset. The same host
application and plugin contracts handle parameter configuration, image acquisition, analysis, and report generation across all models.
Figure~\ref{fig:plugin-ui} shows the dark-current plugin interfaces for the CTS (left) and MCT (right) configurations. The CTS panel provides flexible parameter control for research use, while the MCT panel adopts a simplified layout for batch testing. Both plugins share the same workflow stages of configuration, acquisition, analysis, and status reporting, and support multiple detector types. Table~\ref{tab:detector-coverage} summarizes the implementation coverage for this evaluation; the transport interface is omitted from the table because all three cameras support both USB and fiber operation.

\begin{table}[t]
\centering
\caption{Camera and detector models exercised with the \pxv{} implementation.}
\label{tab:detector-coverage}
\small
\begin{tabular}{>{\raggedright\arraybackslash}p{0.20\linewidth}>{\raggedright\arraybackslash}p{0.22\linewidth}>{\raggedright\arraybackslash}p{0.48\linewidth}}
\toprule
Camera & Detector & Evaluation performed \\
\midrule
VPX1081 (CMOS) & Gpixel GSENSE1081BSI (CMOS) & Full integration, workflow execution, and report generation \\
MCT20 (HgCdTe/MCT) & HgCdTe/MCT detector & Integration and workflow execution \\
CCD4720 (CCD) & Teledyne e2v CCD47-20 (CCD) & Integration and workflow execution \\
\bottomrule
\end{tabular}
\end{table}

\subsection{Workflow execution}

In the legacy workflow, operators relied on Nezha acquisition software or camera-specific Python scripts. The process was fragmented: model-dependent commands, serial temperature control, acquisition-sequence entry, image and frame-count inspection, data assembly, FITS conversion, metadata entry, analysis, and reporting were performed as separate steps with no common execution context. In contrast, a \pxv{} run centralizes these operations through a plugin panel. The user selects the test, enters the scientific parameters, starts the run, and confirms a temperature point when prompted. The platform then automatically handles camera enumeration, preset loading, device-identity capture, exposure-sequence execution, FITS and metadata writing, frame-count validation, analysis, report generation, and index updates. If an acquisition is incomplete, the system reports the issue to the operator rather than automatically reacquiring. The SDK records temperature readings, and the operator must still confirm at each requested temperature point.

\begin{figure}[ht]
    \centering
    \includegraphics[width=\linewidth]{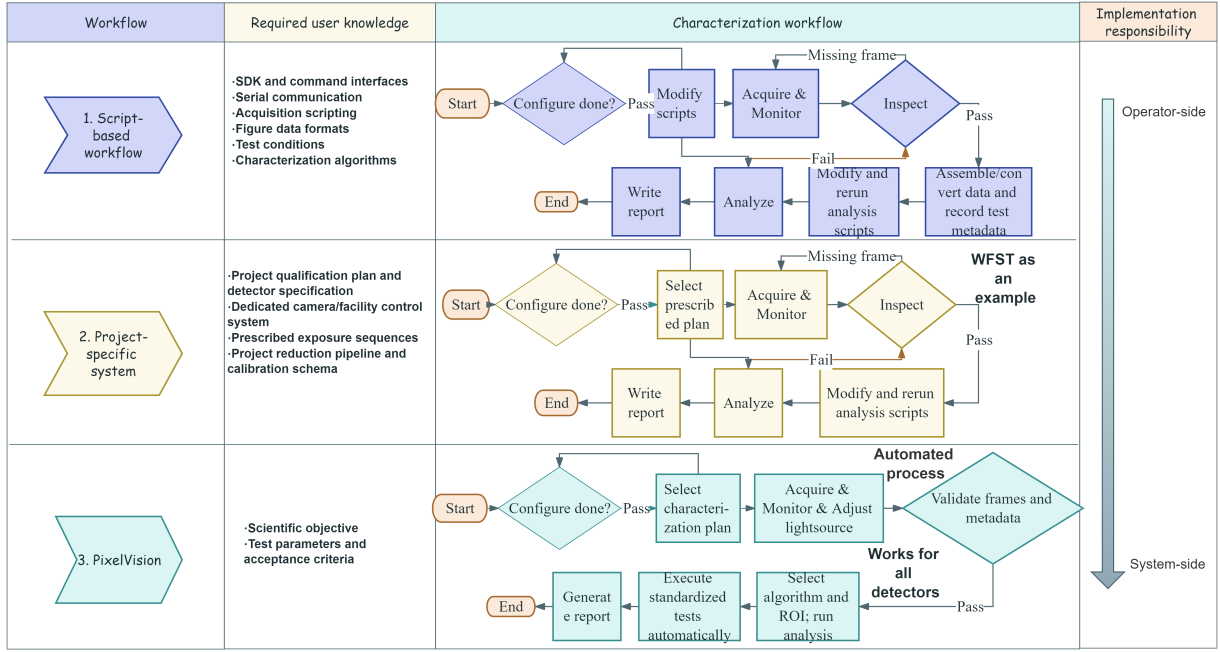}
    \caption{Comparison of knowledge requirements, user operations, and system automation across three detector characterization workflows.}
    \label{fig:workflow-comparison}
\end{figure}

Figure~\ref{fig:workflow-comparison} contrasts the knowledge required, the user-performed operations, and the system processing in the legacy workflow, representative project-specific detector-test systems, and \pxv{}. The project-specific column summarizes common workflow components reported for Euclid NISP, LSST Camera, Roman H4RG, and configurable laboratory testbeds \cite{secroun2016euclidh2rg,snyder2020lsst,mosby2020romanh4rg,khan2025ccd,shapiro2019ppl}. The comparison shows which tasks the operator performs and which the system handles.

\subsection{VPX1081 CMOS end-to-end case}

\begin{figure}[t]
\centering
\includegraphics[width=0.86\linewidth]{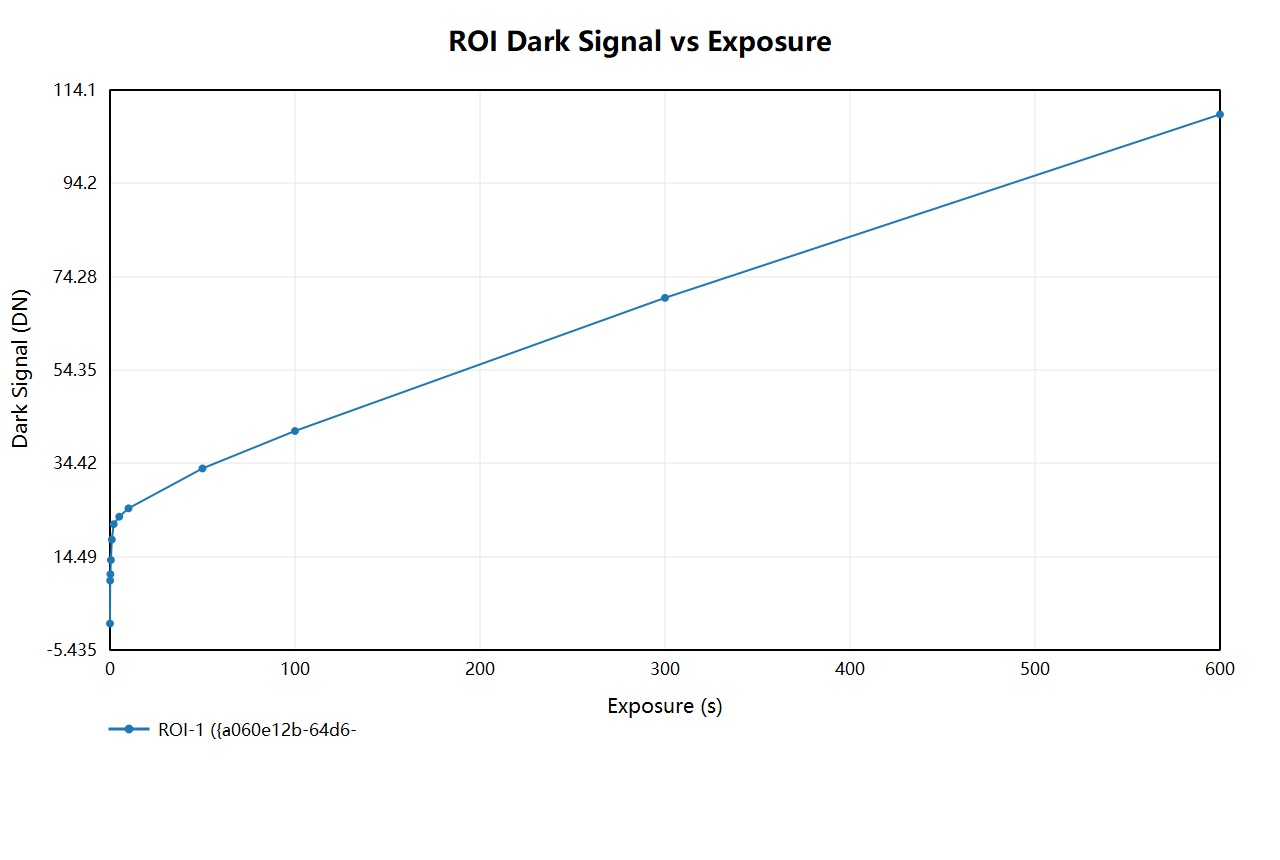}
\caption{ROI dark signal versus exposure time in the platform-generated VPX1081 dark-current report. The curve is stored with the acquisition parameters and metric records in the run archive.}
\label{fig:dark-current-case}
\end{figure}

Platform-generated reports for the VPX1081 camera provide an end-to-end implementation case covering photon transfer, dark current, spatial nonuniformity, and bias-frame noise \cite{rosenberger2016emva1288}. The dark-current run acquired ten bias frames and ten dark frames at each of eleven exposure times:
\[
\{0.1,0.2,0.5,1,2,5,10,50,100,300,600\}\ \mathrm{s}.
\]
The plugin analyzed a $500\times500$-pixel ROI at $(x,y)=(0,3805)$. The run archive contains the FITS frames, machine-readable metric and ROI summaries, time-series and histogram artifacts, and the generated PDF report. For this ROI, the report gives a dark current of $0.1812$~DN/s, a dark-current nonuniformity (DCNU) of $1.6178$~DN/s, and 665 hot/bad pixels (classified using the platform's default threshold) \cite{widenhorn2002temperature,rosenberger2016emva1288}. Figure~\ref{fig:dark-current-case} shows the corresponding ROI dark-signal curve.

\begin{table}[h]
\centering
\caption{Principal outputs from the available VPX1081 platform reports. Values refer to the ROI used in each report; its dimensions can differ among tests.}
\label{tab:uploaded-results}
\small
\begin{tabular}{>{\raggedright\arraybackslash}p{0.19\linewidth}>{\raggedright\arraybackslash}p{0.23\linewidth}>{\raggedright\arraybackslash}p{0.30\linewidth}>{\raggedright\arraybackslash}p{0.16\linewidth}}
\toprule
Test report & Acquisition summary & Principal output & Unit and scope \\
\midrule
Photon transfer & 26 exposure points; 10 frames/point & Gain $1.372$; full-well capacity $81894$; $R^2=0.9999$ & e$^-$/DN and e$^-$ \\
Dark current & 10 bias frames; 11 dark exposure points; 10 frames/point & Dark current $0.1812$; DCNU $1.6178$; 665 hot/bad pixels & DN/s and count/$500\times500$ ROI \\
Spatial nonuniformity & 200 flat frames & PRNU $0.019451$ & ratio/ROI \\
Bias-frame noise & 100 bias frames & Median temporal noise $5.1300$ & e$^-$/$500\times500$ ROI \\
\bottomrule
\end{tabular}
\end{table}

Table~\ref{tab:uploaded-results} lists the main outputs of the four reports. The photon-transfer analysis, based on 26 exposure points, yields a conversion gain of $1.372$~e$^-$/DN, a full-well capacity of $81894$~e$^-$, and $R^2=0.9999$ for the selected ROI \cite{janesick1987photon,rosenberger2016emva1288}. The spatial-nonuniformity measurement reports a PRNU of $0.019451$ from 200 flat frames \cite{rosenberger2016emva1288}. The bias-frame-noise measurement gives a median temporal noise of $5.1300$~e$^-$ for a $500\times500$ ROI, obtained by applying the $1.372$~e$^-$/DN conversion gain from the photon-transfer analysis to the measured $3.7391$~DN \cite{yu2018kasi}. These values come from the implemented workflows and therefore demonstrate the platform's end-to-end execution and artifact-generation capability.

\section{Conclusion}

This work demonstrates that a common software architecture can characterize heterogeneous scientific cameras while leaving the host application unchanged across detector models. In \pxv{}, hardware access is separated from characterization logic through VPXSDK, declarative presets, and runtime plugins. At the same time, the run archive retains acquired data, operating conditions, configurations, analysis products, and reports for each test. The platform was exercised with VPX1081 (CMOS), VPXIR-MCT20 (HgCdTe/MCT), and CCD4720 (CCD) cameras, and the VPX1081 experiments verified an end-to-end workflow from acquisition to characterization reports. ChatPixel extends the platform with read-only, source-linked assistance for operation and diagnostics, while withholding operational answers when supporting evidence is unavailable. These results show the feasibility of using the same host, workflow, and data-management model across different detector technologies. Future work will extend detector coverage and evaluate ChatPixel over a broader range of operational and diagnostic cases.

\section{Disclosures}
The authors declare there are no financial interests, commercial affiliations, or other potential conflicts of interest that have influenced the objectivity of this research or the writing of this paper.

\section{Code and Data Availability}
No code is publicly available with this manuscript due to proprietary and project-related restrictions. The data supporting the findings of this study are also not publicly available due to proprietary and project-related restrictions, but may be available from the corresponding author upon reasonable request and subject to applicable restrictions.

\section{Acknowledgments}
This work was supported in part by the Fundamental Research Funds for the Central Universities under Grant WK2360000016, Grant YD2030000601, Grant YD2030000602, and Grant YD2030000604; in part by the ET space mission, which is funded by China's Space Origins Exploration Program (GJ11030211); in part by the research funds of the National Key Lab of Deep Space Exploration No.NKDSEL2024009, in part by the Strategic Priority Program on Space Science, the Chinese Academy of Sciences (Grant No. XDA15020605), in part by Research Funds of the State Key Laboratory of Particle Detection and Electronics (SKLPDE-ZZ-202325 and SKLPDE-KF-202314), in
part by the Cyrus Chun Ying Tang Foundation, and in part by Frontier Scientific Research Program of Deep Space Exploration Laboratory (Grant No. 2022-QYKYJH-HXYF-012).

DeepSeek was used during the preparation of this work to assist with debugging a limited portion of the software and to edit the manuscript for language and grammar. Prompts included requests to help identify possible causes of software errors, suggest debugging approaches, and improve the grammar, clarity, and readability of the manuscript without altering its technical content. The authors reviewed and verified all AI-assisted debugging suggestions and textual edits before use.

\bibliography{references}

@article{wang2023wfst,
  author = {Wang, T. and Liu, G. and Cai, Z. and et al.},
  title = {Science with the 2.5-meter {Wide Field Survey Telescope} ({WFST})},
  journal = {Sci. China Phys. Mech. Astron.},
  volume = {66},
  pages = {109512},
  year = {2023},
  doi = {10.1007/s11433-023-2197-5},
}

@article{ge2022earth20,
  author = {Ge, Jian and Zhang, Hui and Deng, Hongping and Howell, Steve B. and {ET team}},
  title = {The {ET} mission to search for {Earth 2.0s}},
  journal = {Innovation},
  volume = {3},
  number = {4},
  pages = {100271},
  year = {2022},
  doi = {10.1016/j.xinn.2022.100271},
}

@article{healey1994radiometric,
  author = {Healey, G. E. and Kondepudy, R.},
  title = {Radiometric {CCD} camera calibration and noise estimation},
  journal = {IEEE Trans. Pattern Anal. Mach. Intell.},
  volume = {16},
  number = {3},
  pages = {267--276},
  year = {1994},
  doi = {10.1109/34.276126},
}

@article{simmhan2005provenance,
  author = {Simmhan, Yogesh L. and Plale, Beth and Gannon, Dennis},
  title = {A survey of data provenance in e-science},
  journal = {ACM SIGMOD Rec.},
  volume = {34},
  number = {3},
  pages = {31--36},
  year = {2005},
  doi = {10.1145/1084805.1084812},

}

@article{khan2025ccd,
  author = {Khan, Aafaque R. and Hamden, Erika and Kyne, Gillian and Jewell, April D. and Hennessy, John and Nikzad, Shouleh and Jones, Todd and Kerkeser, Nazende Ipek and Parker, Brock and Picouet, Vincent and Lin, Zeren and Vider, Jacob and Verma, Shashank and Bradley, Harrison and Beaty, David and Gacon, Frank and Ford, John and West, Grant and Arbo, Paul},
  title = {Comprehensive detector test platform with precision thermal control for noise characterization of charge-coupled devices},
  journal = {J. Astron. Telesc. Instrum. Syst.},
  volume = {11},
  number = {4},
  pages = {042204},
  year = {2025},
  doi = {10.1117/1.JATIS.11.4.042204},

}

@article{shapiro2019ppl,
  author = {Shapiro, Charles and Smith, Roger and Huff, Eric and Plazas, Andr{\'e}s A. and Rhodes, Jason and Fucik, Jason and Goodsall, Tim and Massey, Richard and Rowe, Barnaby and Seshadri, Suresh},
  title = {{Precision Projector Laboratory}: detector characterization with an astronomical emulation testbed},
  journal = {J. Astron. Telesc. Instrum. Syst.},
  volume = {5},
  number = {4},
  pages = {041503},
  year = {2019},
  doi = {10.1117/1.JATIS.5.4.041503},

}

@inproceedings{snyder2020lsst,
  author = {Snyder, Adam and Barrau, Aur{\'e}lien and Bradshaw, Andrew and Bowdish, Boyd W. and Chiang, James and Combet, Celine and Digel, Seth and Dubois, Richard and Eraud, Ludovic and Juramy, Claire G. and Lage, Craig S. and Lange, Travis and Migliore, Myriam and Nomerotski, Andrei and O'Connor, Paul and Park, HyeYun and Rasmussen, Andrew and Reil, Kevin and Roodman, Aaron and Shestakov, Adrian and Utsumi, Yousuke and Wood, Duncan},
  title = {Laboratory measurements of instrumental signatures of the {LSST Camera} focal plane},
  booktitle = {X-Ray, Optical, and Infrared Detectors for Astronomy IX},
  series = {Proc. SPIE},
  volume = {11454},
  pages = {1145439},
  year = {2020},
  doi = {10.1117/12.2562915},

}

@article{pence2010fits,
  author = {Pence, W. D. and Chiappetti, L. and Page, C. G. and Shaw, R. A. and Stobie, E.},
  title = {Definition of the {Flexible Image Transport System} ({FITS}), version 3.0},
  journal = {A\&A},
  volume = {524},
  pages = {A42},
  year = {2010},
  doi = {10.1051/0004-6361/201015362},

}

@article{rieke2015miri,
  author = {Rieke, G. H. and Ressler, M. E. and Morrison, Jane E. and Bergeron, L. and Bouchet, Patrice and Garc{\'i}a-Mar{\'i}n, Macarena and Greene, T. P. and Regan, M. W. and Sukhatme, K. G. and Walker, Helen},
  title = {The {Mid-Infrared Instrument} for the {James Webb Space Telescope}, {VII}: The {MIRI} Detectors},
  journal = {PASP},
  volume = {127},
  number = {953},
  pages = {665--674},
  year = {2015},
  doi = {10.1086/682257},

}

@inproceedings{schindler2014ingaasbench,
  author = {Schindler, Karsten and Wolf, J{\"u}rgen and Krabbe, Alfred},
  title = {Characterization of {InGaAs}-based cameras for astronomical applications using a new {VIS-NIR-SWIR} detector test bench},
  booktitle = {Ground-based and Airborne Telescopes V},
  editor = {Stepp, Larry M. and Gilmozzi, Roberto and Hall, Helen J.},
  series = {Proc. SPIE},
  volume = {9145},
  pages = {91450X},
  year = {2014},
  publisher = {SPIE},
  doi = {10.1117/12.2057052},

}

@inproceedings{secroun2016euclidh2rg,
  author = {Secroun, Aur{\'e}lia and Serra, Benoit and Cl{\'e}mens, Jean Claude and Legras, Romain and Lagier, Philippe and Niclas, Mathieu and Caillat, Laurence and Gillard, William and Tilquin, Andr{\'e} and Ealet, Anne and Barbier, R{\'e}mi and Ferriol, Sylvain and Kubik, Bogna and Smadja, G{\'e}rard and Prieto, Eric and Maciaszek, Thierry and Norup Sorensen, Anton},
  title = {Characterization of {H2RG IR} detectors for the {Euclid NISP} instrument},
  booktitle = {High Energy, Optical, and Infrared Detectors for Astronomy VII},
  editor = {Holland, Andrew D. and Beletic, James W.},
  series = {Proc. SPIE},
  volume = {9915},
  pages = {99151Y},
  year = {2016},
  publisher = {SPIE},
  doi = {10.1117/12.2232070},

}

@inproceedings{waczynski2016euclid,
  author = {Waczynski, A. and Barbier, R. and Cagiano, S. and Chen, J. and Cheung, S. and Cho, H. and Cillis, A. and Cl{\'e}mens, J.-C. and Dawson, O. and Delo, G. and Farris, M. and Feizi, A. and Foltz, R. and Hickey, M. and Holmes, W. and Hwang, T. and Israelsson, U. and Jhabvala, M. and Kahle, D. and Kan, Em. and Kan, Er. and Loose, M. and Lotkin, G. and Miko, L. and Nguyen, L. and Piquette, E. and Powers, T. and Pravdo, S. and Runkle, A. and Seiffert, M. and Strada, P. and Tucker, C. and Turck, K. and Wang, F. and Weber, C. and Williams, J.},
  title = {Performance overview of the {Euclid} infrared focal plane detector subsystems},
  booktitle = {High Energy, Optical, and Infrared Detectors for Astronomy VII},
  series = {Proc. SPIE},
  volume = {9915},
  pages = {991511},
  year = {2016},
  doi = {10.1117/12.2231641},

}

@article{mosby2020romanh4rg,
  author = {Mosby, Gregory and Rauscher, Bernard J. and Bennett, Chris and Cheng, Edward S. and Cheung, Stephanie and Cillis, Analia and Content, David and Cottingham, Dave and Foltz, Roger and Gygax, John and Hill, Robert J. and Kruk, Jeffrey W. and Mah, Jon and Meier, Lane and Merchant, Chris and Miko, Laddawan and Piquette, Eric C. and Waczynski, Augustyn and Wen, Yiting},
  title = {Properties and characteristics of the {Nancy Grace Roman Space Telescope H4RG-10} detectors},
  journal = {J. Astron. Telesc. Instrum. Syst.},
  volume = {6},
  number = {4},
  pages = {046001},
  year = {2020},
  doi = {10.1117/1.JATIS.6.4.046001},

}

@article{dalesio1994epics,
  author = {Dalesio, Leo R. and Hill, Jeffrey O. and Kraimer, Martin and Lewis, Stephen and Murray, Douglas and Hunt, Stephan and Watson, William and Clausen, Matthias and Dalesio, John},
  title = {The Experimental Physics and Industrial Control System Architecture: Past, Present, and Future},
  journal = {Nucl. Instrum. Methods Phys. Res. A},
  volume = {352},
  number = {1--2},
  pages = {179--184},
  year = {1994},
  doi = {10.1016/0168-9002(94)91493-1},

}

@inproceedings{chiozzi2004acs,
  author = {Chiozzi, Gianluca and Jeram, Bogdan and Sommer, Heiko and Caproni, Alessandro and Plesko, Mark and Sekoranja, Matej and Zagar, Klemen and Fugate, David W. and Di Marcantonio, Paolo and Cirami, Roberto},
  title = {The {ALMA Common Software}: A Developer-Friendly {CORBA}-Based Framework},
  booktitle = {Advanced Software, Control, and Communication Systems for Astronomy},
  series = {Proc. SPIE},
  volume = {5496},
  pages = {205--218},
  year = {2004},
  doi = {10.1117/12.551943},

}

@article{kubanek2010rts2,
  author = {Kub{\'a}nek, Petr},
  title = {{RTS2}---The Remote Telescope System},
  journal = {Adv. Astron.},
  volume = {2010},
  pages = {902484},
  year = {2010},
  doi = {10.1155/2010/902484},

}

@article{husser2022pyobs,
  author = {Husser, Tim-Oliver and Hessman, Frederic V. and Martens, Sven and Masur, Tilman and Royen, Karl and Sch{\"a}fer, Sebastian},
  title = {pyobs---An Observatory Control System for Robotic Telescopes},
  journal = {Front. Astron. Space Sci.},
  volume = {9},
  pages = {891486},
  year = {2022},
  doi = {10.3389/fspas.2022.891486},

}

@article{allan2019bluesky,
  author = {Allan, Daniel and Caswell, Thomas and Campbell, Stuart and Rakitin, Maksim},
  title = {Bluesky's Ahead: A Multi-Facility Collaboration for an \textit{a la Carte} Software Project for Data Acquisition and Management},
  journal = {Synchrotron Radiat. News},
  volume = {32},
  number = {3},
  pages = {19--22},
  year = {2019},
  doi = {10.1080/08940886.2019.1608121},

}

@inproceedings{lewis2020rag,
  author = {Lewis, Patrick and Perez, Ethan and Piktus, Aleksandra and Petroni, Fabio and Karpukhin, Vladimir and Goyal, Naman and K{\"u}ttler, Heinrich and Lewis, Mike and Yih, Wen-tau and Rockt{\"a}schel, Tim and Riedel, Sebastian and Kiela, Douwe},
  title = {Retrieval-Augmented Generation for Knowledge-Intensive {NLP} Tasks},
  booktitle = {Advances in Neural Information Processing Systems},
  volume = {33},
  pages = {9459--9474},
  year = {2020},
  url = {https://proceedings.neurips.cc/paper/2020/hash/6b493230205f780e1bc26945df7481e5-Abstract.html},

}

@article{robertson2009bm25,
  author = {Robertson, Stephen E. and Zaragoza, Hugo},
  title = {The Probabilistic Relevance Framework: {BM25} and Beyond},
  journal = {Found. Trends Inf. Retr.},
  volume = {3},
  number = {4},
  pages = {333--389},
  year = {2009},
  doi = {10.1561/1500000019},

}

@inproceedings{yu2020queryrewriting,
  author = {Yu, Shi and Liu, Jiahua and Yang, Jingqin and Xiong, Chenyan and Bennett, Paul N. and Gao, Jianfeng and Liu, Zhiyuan},
  title = {Few-Shot Generative Conversational Query Rewriting},
  booktitle = {Proceedings of the 43rd International ACM SIGIR Conference on Research and Development in Information Retrieval},
  pages = {1933--1936},
  year = {2020},
  doi = {10.1145/3397271.3401323},

}

@inproceedings{gao2023citations,
  author = {Gao, Tianyu and Yen, Howard and Yu, Jiatong and Chen, Danqi},
  title = {Enabling Large Language Models to Generate Text with Citations},
  booktitle = {Proceedings of the 2023 Conference on Empirical Methods in Natural Language Processing},
  pages = {6465--6488},
  address = {Singapore},
  publisher = {Association for Computational Linguistics},
  year = {2023},
  doi = {10.18653/v1/2023.emnlp-main.398},

}

@article{ji2023hallucination,
  author = {Ji, Ziwei and Lee, Nayeon and Frieske, Rita and Yu, Tiezheng and Su, Dan and Xu, Yan and Ishii, Etsuko and Bang, Ye Jin and Madotto, Andrea and Fung, Pascale},
  title = {Survey of Hallucination in Natural Language Generation},
  journal = {ACM Comput. Surv.},
  volume = {55},
  number = {12},
  pages = {1--38},
  year = {2023},
  doi = {10.1145/3571730},

}

@inproceedings{rajpurkar2018unanswerable,
  author = {Rajpurkar, Pranav and Jia, Robin and Liang, Percy},
  title = {Know What You Don't Know: Unanswerable Questions for {SQuAD}},
  booktitle = {Proceedings of the 56th Annual Meeting of the Association for Computational Linguistics (Volume 2: Short Papers)},
  pages = {784--789},
  address = {Melbourne, Australia},
  publisher = {Association for Computational Linguistics},
  year = {2018},
  doi = {10.18653/v1/P18-2124},

}

@article{yu2018kasi,
  author  = {Yu, Young Sam and Kim, Jinsol and Park, Chan
             and Jeong, Woong-Seob and Kim, Minjin
             and Choi, Seonghwan and Park, Sung-Joon},
  title   = {Evaluation of the {KASI Detector Performance Test System}
             Using an {Andor iKon M} {CCD} Camera},
  journal = {J. Astron. Space Sci.},
  volume  = {35},
  number  = {3},
  pages   = {201--210},
  year    = {2018},
  doi     = {10.5140/JASS.2018.35.3.201},

}

@article{janesick1987photon,
  author  = {Janesick, James R. and Klaasen, Kenneth P. and Elliott, Tom},
  title   = {Charge-Coupled-Device Charge-Collection Efficiency
             and the Photon-Transfer Technique},
  journal = {Opt. Eng.},
  volume  = {26},
  number  = {10},
  pages   = {972--980},
  year    = {1987},
  doi     = {10.1117/12.7974183},

}

@inproceedings{widenhorn2002temperature,
  author    = {Widenhorn, Ralf and Blouke, Morley M. and Weber, Alexander
               and Rest, Armin and Bodegom, Erik},
  title     = {Temperature Dependence of Dark Current in a {CCD}},
  booktitle = {Sensors and Camera Systems for Scientific, Industrial,
               and Digital Photography Applications III},
  series    = {Proc. SPIE},
  volume    = {4669},
  pages     = {193--201},
  year      = {2002},
  doi       = {10.1117/12.463446},

}

@article{kazic2015provenance,
  author  = {Kazic, Toni},
  title   = {Ten Simple Rules for Experiments' Provenance},
  journal = {PLoS Comput. Biol.},
  volume  = {11},
  number  = {10},
  pages   = {e1004384},
  year    = {2015},
  doi     = {10.1371/journal.pcbi.1004384},

}

@inproceedings{areadetectorReference,
  author    = {Rivers, Mark L.},
  title     = {{areaDetector}: {EPICS} Software for 2-{D} Detectors},
  booktitle = {Proceedings of the 16th International Conference on
               Accelerator and Large Experimental Control Systems
               (ICALEPCS 2017)},
  year      = {2018},
  pages     = {1245--1251},
  address   = {Barcelona, Spain},
  publisher = {JACoW},
  doi       = {10.18429/JACoW-ICALEPCS2017-THDPL03},
  url       = {https://doi.org/10.18429/JACoW-ICALEPCS2017-THDPL03},
}

@inproceedings{limaReference,
  author    = {Petitdemange, Samuel and
               Claustre, Laurent and
               Henry, Anthony and
               Homs Regojo, Ricardo and
               Homs, Alejandro and
               Langlois, Fr{\'e}d{\'e}ric and
               Mant, Geoffrey R. and
               Naudet, Didier and
               Noureddine, Ahmed and
               Papillon, Eric},
  title     = {{LIMA}: Library for {IM}age Acquisition, a Worldwide
               Project for 2-{D} Detector Control},
  booktitle = {Proceedings of the 16th International Conference on
               Accelerator and Large Experimental Control Systems
               (ICALEPCS 2017)},
  year      = {2018},
  pages     = {886--890},
  address   = {Barcelona, Spain},
  publisher = {JACoW},
  doi       = {10.18429/JACoW-ICALEPCS2017-TUPHA194},
  url       = {https://doi.org/10.18429/JACoW-ICALEPCS2017-TUPHA194},
}

@article{stuurman2010micromanager,
  author  = {Edelstein, Arthur and Amodaj, Nenad and Hoover, Karl
             and Vale, Ronald and Stuurman, Nico},
  title   = {Computer Control of Microscopes Using {Micro-Manager}},
  journal = {Curr. Protoc. Mol. Biol.},
  year    = {2010},
  volume  = {92},
  number  = {1},
  pages   = {14.20.1--14.20.17},
  doi     = {10.1002/0471142727.mb1420s92},
}

@inproceedings{prouza2010rts2ccd,
  author    = {Prouza, Michael and Kub{\'a}nek, Petr and O'Connor, Paul
               and Kotov, Ivan and Frank, James and Antilogus, Pierre},
  title     = {Software for Automated Testing and Characterization of
               {CCD}s for the {Large Synoptic Survey Telescope} ({LSST})},
  booktitle = {Software and Cyberinfrastructure for Astronomy},
  series    = {Proc. SPIE},
  volume    = {7740},
  pages     = {77401Z},
  year      = {2010},
  publisher = {SPIE},
  doi       = {10.1117/12.857340},
}

@inproceedings{gotz2003tango,
  author = {G{\"o}tz, A. and Taurel, E. and Pons, J. L. and Verdier, P. and Chaize, J. M. and Meyer, J. and Poncet, F. and Heunen, G. and G{\"o}tz, E. and Buteau, A. and Leclercq, N. and Ounsy, M.},
  title = {{TANGO}: A {CORBA}-Based Control System},
  booktitle = {Proceedings of the 9th International Conference on Accelerator and Large Experimental Physics Control Systems (ICALEPCS 2003)},
  pages = {220--222},
  address = {Gyeongju, Korea},
  year = {2003}
}

@article{hoffmann2022indi,
  author = {Hoffmann, Tobias and Gehlen, Matti and Plaggenborg, Thorsten and Drolshagen, Gerhard and Ott, Theresa and Kunz, Jutta and Santana-Ros, Toni and Gedek, Marcin and Reszelewski, Rafa{\l} and {\.Z}o{\l}nowski, Micha{\l} and Poppe, Bj{\"o}rn},
  title = {Robotic Observation Pipeline for Small Bodies in the Solar System Based on Open-Source Software and Commercially Available Telescope Hardware},
  journal = {Front. Astron. Space Sci.},
  volume = {9},
  pages = {895732},
  year = {2022},
  doi = {10.3389/fspas.2022.895732},
}

@article{rosenberger2016emva1288,
  author = {Rosenberger, Maik and Zhang, Chen and Votyakov, Pavel and Prei{\ss}ler, Marc and Celestre, Rafael and Notni, Gunther},
  title = {{EMVA 1288} Camera Characterisation and the Influences of Radiometric Camera Characteristics on Geometric Measurements},
  journal = {Acta IMEKO},
  volume = {5},
  number = {4},
  pages = {81--87},
  year = {2016},
  doi = {10.21014/acta_imeko.v5i4.356},

}

@article{moreau2015prov,
  author = {Moreau, Luc and Groth, Paul and Cheney, James and Lebo, Timothy and Miles, Simon},
  title = {The Rationale of {PROV}},
  journal = {Web Semant.},
  volume = {35},
  pages = {235--257},
  year = {2015},
  doi = {10.1016/j.websem.2015.04.001},
}

@inproceedings{mulholland2020genicam,
  author = {Mulholland, K. F. and Knudstrup, J. and Pellegrin, F.},
  title = {A New Communication Interface for the {European Southern Observatory (ESO)}'s {Very Large Telescope} Technical Detector Control System Using {Aravis}, an Open-Source Library for {GenICam} Cameras},
  booktitle = {Proceedings of the 17th International Conference on Accelerator and Large Experimental Physics Control Systems (ICALEPCS 2019)},
  pages = {444--447},
  publisher = {JACoW Publishing},
  address = {Geneva, Switzerland},
  year = {2020},
  doi = {10.18429/JACoW-ICALEPCS2019-MOPHA098},
}

@article{zhao2023bluesky,
  author = {Zhao, Ying and Hu, Chun and Wang, Chunpeng and Cao, Jiefeng and Zhang, Zhaohong},
  title = {Data Acquisition System Based on the {Bluesky} Suite in the {Shanghai Synchrotron Radiation Facility}},
  journal = {Appl. Sci.},
  volume = {13},
  number = {10},
  pages = {5829},
  year = {2023},
  doi = {10.3390/app13105829},
}
\bibliographystyle{spiejour}

\end{spacing}
\end{document}